\documentstyle[twocolumn,prl,aps,epsfig]{revtex}

\def\cb{\bar{\chi}}
\def\cbe{\bar{\chi}_{\rm eff}}

\begin{document}
\draft

\twocolumn[\hsize\textwidth\columnwidth\hsize\csname
@twocolumnfalse\endcsname

\begin{flushright}
ANL-HEP-PR-00-116\\
\end{flushright}

\preprint{ANL-HEP-PR-00-116}
\title{Low-Energy Supersymmetry and the Tevatron Bottom-Quark Cross
Section}

\author{E.~L.~Berger$^1$, B.~W.~Harris$^1$, D.~E.~Kaplan$^{1,2}$,
Z.~Sullivan$^1$, T.~M.~P.~Tait$^1$ and C.~E.~M.~Wagner$^{1,2}$\\}
\address{
$^1$ High Energy Physics Division, Argonne National Laboratory, Argonne,
IL 60439 \\
$^2$ Enrico Fermi Institute, University of Chicago, Chicago,
IL 60637 \\}
\date{November 30, 2000}
\maketitle
\begin{abstract}
A long-standing discrepancy between the bottom-quark production cross
section and predictions of perturbative quantum chromodynamics is
addressed.  We show that pair production of light gluinos, of mass 12 to 16
GeV, with two-body decays into bottom quarks and light bottom squarks,
yields a bottom-quark production rate in agreement with hadron collider
data.  We examine constraints on this scenario from low-energy data and
make predictions that may be tested at the next run of the Fermilab Tevatron
Collider.
\end{abstract}

\vskip1.0pc]


The measured cross section for bottom-quark production at hadron collider
energies exceeds the expectations of next-to-leading order calculations in
perturbative quantum chromodynamics (QCD) by about a factor of 2
\cite{expxsec}.  The next-to-leading order (NLO) corrections are large, and
it is not excluded that higher order effects in production or fragmentation
may resolve the discrepancy.  However, this long-standing discrepancy has
so far resisted satisfactory resolution within the standard model (SM) of
particle physics~\cite{qcdrev}.  The disagreement is surprising because the
relatively large mass of the bottom quark sets a hard scattering scale at
which perturbative QCD computations of other processes are generally
successful.  In this Letter we explore an explanation of the discrepancy
within the context of the minimal supersymmetric standard model
(MSSM)~\cite{mssm}.  We postulate the existence of a relatively light
gluino $\tilde g$ (mass $\simeq 12$--16 GeV) that decays into a bottom
quark and a light bottom squark $\tilde b$ (mass $\simeq 2$--5.5 GeV).  The
$\tilde g$ and the $\tilde b$ are the spin-1/2 and spin-0 supersymmetric
partners of the gluon ($g$) and bottom quark ($b$).  In our scenario the
$\tilde b$ is either long-lived or decays hadronically.  We obtain good
agreement with hadron collider rates of bottom-quark production.  Several
predictions are made that can be tested readily with forthcoming data from
run II of the Fermilab Tevatron collider.

Our assumptions are consistent with all experimental constraints on the
masses and couplings of supersymmetric
particles~\cite{barate,uaone,CHWW,Nappi,besii,CELLO,CLEO,PLEHN,baeretal}.
An analysis of four-jet data by the ALEPH Collaboration disfavors a $\tilde
g$ with mass $m_{\tilde g} < 6.3$ GeV \cite{barate} but does not cover the
mass range considered in this Letter.  An analysis by the UA1 Collaboration
\cite{uaone} of the mass range $4<m_{\tilde g}<53$ GeV does not apply to
our scenario because they assume that the gluino decays into two quarks
plus large missing energy.  However, a new comparison to the UA1 data
\cite{qcdrev} does see an excess in the $b$-quark cross section, as would
be expected in our model.

If the light $\tilde{b}$ is an appropriate admixture of left-handed and
right-handed bottom squarks, its tree-level coupling to the neutral gauge
boson $Z$ can be small, leading to good agreement with the $Z$-peak
observables~\cite{CHWW}.  Bottom squarks make a tiny contribution to the
inclusive cross section for $e^+ e^- \rightarrow$ hadrons, in comparison to
the contributions from quark production, and $\tilde{b} \bar{\tilde{b}}$
resonances are likely to be impossible to extract from backgrounds
\cite{Nappi,besii}.  One can study the angular distribution of hadronic
jets produced in $e^+ e^-$ annihilation in order to bound the contribution
of scalar-quark production.  Spin-1/2 quarks and spin-0 squarks emerge with
different distributions, $(1 \pm {\rm cos}^2 \theta)$, respectively.  We
refit the angular distribution measured by the CELLO
Collaboration~\cite{CELLO} and find it is consistent with the production
of a single pair of charge-1/3 squarks along with five flavors of
quark-antiquark pairs.  The exclusion by the CLEO Collaboration~\cite{CLEO}
of a $\tilde b$ with mass 3.5 to 4.5 GeV does not apply since that analysis
focuses on the decays $\tilde b \rightarrow c \em{l} \tilde \nu$ and
$\tilde b \rightarrow c {\em l}$ \cite{PLEHN}.  Thus, measurements at $e^+
e^-$ colliders do not significantly constrain $\tilde b$ masses in the
region of interest.

A long-lived $\tilde b$ is not excluded by conventional searches at hadron
and lepton colliders, but an analysis similar to that of
Ref.~\cite{baeretal} should be done to verify that there are no additional
constraints on the allowed range of $\tilde b$ masses and lifetimes.
Alternately, the $\tilde b$ could decay hadronically via a baryon-number-
and $R$-parity-violating interaction into soft light hadrons which will
fall within the cone of the $b$ jet.  The main constraint is that the
$\tilde b$ does not decay into hard leptons or leave a large missing
energy.  The $\tilde b$ and $\tilde g$ masses we consider are also
compatible with theoretical constraints which require the absence of color
and charge breaking minima in the scalar potential \cite{dreiner}.

Because the excess production rate is observed in all bottom-quark decay
channels, an explanation in terms of ``new physics'' is guided towards
hypothesized new particles that decay either like bottom quarks or directly
to bottom quarks.  The former is difficult to implement successfully in the
MSSM, as explained in the discussion of alternative scenarios at the end of
the Letter.  We adopt the latter.  In our scenario, light gluinos are
produced in pairs via standard QCD subprocesses, dominantly $g + g
\rightarrow \tilde g + \tilde g$ at Tevatron energies.  The $\tilde g$ has
a strong color coupling to $b$'s and $\tilde b$'s and, as long as its mass
satisfies $m_{\tilde g} > m_b + m_{\tilde b}$, the $\tilde g$ decays
promptly to $b + \tilde b$.  The magnitude of the $b$ cross section, the
shape of the $b$'s transverse momentum $p_{Tb}$ distribution, and the CDF
measurement~\cite{cdfmix} of $B^0 - \bar B^0$ mixing are three features of
the data that help to establish the preferred masses of the $\tilde g$ and
$\tilde b$.

Including contributions from both $q + \bar q \rightarrow \tilde g + \tilde
g$ and $g + g \rightarrow \tilde g + \tilde g$~\cite{dawson}, we show in
Fig.~1 the integrated $p_{Tb}$ distribution of the $b$ quarks that results
from $\tilde g \rightarrow b + \tilde b$, for $m_{\tilde g} = $14 GeV and
$m_{\tilde b} =$ 3.5 GeV.  The results are compared with the cross section
obtained from NLO perturbative QCD~\cite{NLOQCD} and CTEQ4M parton
distribution functions (PDF's)~\cite{cteq}, with $m_b =$ 4.75 GeV, and a
renormalization and factorization scale $\mu=\sqrt{m_b^2 + p_{Tb}^2}$.
SUSY-QCD corrections to $b \bar{b}$ production are not included as they are
not available and are generally expected to be somewhat smaller than the
standard QCD corrections.  A fully differential NLO calculation of $\tilde
g$-pair production and decay does not exist either.  Therefore, we compute
the $\tilde g$-pair cross section from the leading order (LO) matrix
element with NLO PDF's \cite{cteq}, $\mu=\sqrt{m^2_{\tilde g}+p^2_{T\tilde
g}}$, a two-loop $\alpha_s$, and multiply by 1.9, the ratio of inclusive
NLO to LO cross sections~\cite{prospino}.

A relatively light gluino is necessary in order to obtain a bottom-quark
cross section comparable in magnitude to the pure QCD component.  Values of
$m_{\tilde g} \simeq$ 12--16 GeV are chosen because the resulting $\tilde
g$ decays produce $p_{Tb}$ spectra that are enhanced primarily in the
neighborhood of $p_{Tb}^{\rm min} \simeq m_{\tilde g}$ where the data show
the most prominent enhancement above the QCD expectation.  Larger values of
$m_{\tilde g}$ yield too little cross section to be of interest, and
smaller values produce more cross section than seems tolerated by the ratio
of like-sign to opposite-sign leptons from $b$ decay, as discussed below.
The choice of $m_{\tilde b}$ has an impact on the kinematics of the $b$.
After selections on $p_{Tb}^{\rm min}$, large values of $m_{\tilde b}$
reduce the cross section and, in addition, lead to shapes of the $p_{Tb}$
distribution that agree less well with the data.  The values of $m_{\tilde
b}$ and $m_{\tilde g}$ are correlated; similar results to those shown in
Fig.~1 can be obtained with $m_{\tilde g} \simeq$ 12 GeV, but $m_{\tilde b}
\simeq m_b$.

After the contributions of the NLO QCD and SUSY components are added (solid
curve in Fig.~1), the magnitude of the bottom-quark cross section and the
shape of the integrated $p^{\rm min}_{Tb}$ distribution are described well.
A theoretical uncertainty of roughly $\pm 30\%$ may be assigned to the
final solid curve, associated with variation of the $b$ mass, the scale,
and the parton distributions.
\begin{figure}[tb]
\begin{center}\vspace*{-3ex}
\psfig{figure=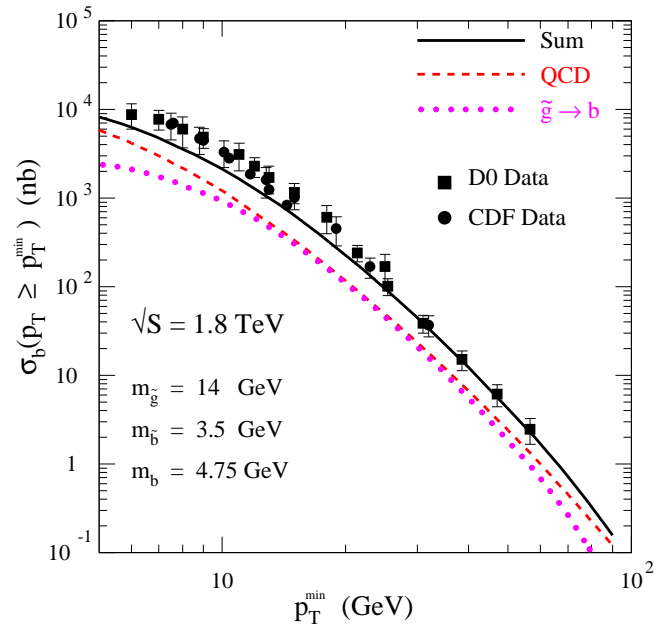,width=3.4in}
\end{center}
\caption{Bottom-quark cross section in $p\bar p$ collisions at $\sqrt{S}
=1.8$ TeV for $p_{Tb}>p_{Tb}^{\rm min}$ with a gluino of mass
$m_{\tilde{g}} = 14$ GeV and a bottom squark of mass $m_{\tilde{b}} = 3.5$
GeV.  The dashed curve is the central value of the NLO QCD prediction. The
dotted curve shows the $p_T$ spectrum of the $b$ from the supersymmetry
(SUSY) processes.  The solid curve is the sum of the QCD and SUSY
components.  Data are from Ref.~\protect\cite{expxsec}.  }
\label{fig:fig1}
\end{figure}
The SUSY process produces bottom quarks in a four-body final state and thus
their momentum correlations are different from those of QCD.  Angular
correlations between muons that arise from decays of $b$'s have been
measured \cite{cdfmix,muonexp}.  Examining the angular correlations between
$b$'s in the SUSY case we find they are nearly indistinguishable from those
of QCD once experimental cuts are applied.

Since the $\tilde g$ is a Majorana particle, its decay yields both quarks
and antiquarks.  Gluino pair production and subsequent decay to $b$'s will
generate $b b$ and $\bar b \bar b$ pairs, as well as the $b \bar b$
final states that appear in QCD production.

We perform an exact matrix-element calculation of the four-body cross
section for like-sign and opposite-sign bottom quarks from $\tilde g$
pair-production and decay.  When a gluino is highly relativistic, its
helicity is nearly the same as its chirality.  Therefore, selection of
$\tilde g$'s whose transverse momentum is greater than their mass will
reduce the number of like-sign $b$'s.  In the limit of either massless
$\tilde{g}$'s or very high $p_T$ cuts, the like-sign to opposite-sign ratio
reduces to $y/(1-y)$ where $y = {1\over 2}\sin^2 2\theta_{\tilde{b}}$ and
$\sin\theta_{\tilde{b}}$ is the left-handed component of the lightest
bottom-squark mass eigenstate.  There is a strong suppression of like-sign
$b$'s if the mixing angle is small.  For the case under consideration, the
mixing of the bottom squark is determined by the condition that the $\tilde
b$ coupling to the $Z$ boson is small~\cite{CHWW}, namely,
$\sin\theta_{\tilde{b}} \simeq 0.38$.  In the intermediate $p_T$ region,
however, the like-sign suppression is reduced.  The cuts chosen in hadron
collider experiments for measurement of the like-sign to opposite-sign muon
ratio result in primarily unpolarized $\tilde g$'s, and, independent of the
$\tilde b$ mixing angle, an equal number of like-sign and opposite-sign
$b$'s is expected at production.

The SUSY mechanism leads to an increase of like-sign leptons in the final
state after semi-leptonic decays of the $b$ and $\bar b$ quarks.  This
increase could be confused with an enhanced rate of $B^0-\bar B^0$ mixing.
Time-integrated mixing analyses of lepton pairs observed at hadron
colliders determine the quantity $\cb$, expressed conventionally as $\cb =
f_d \chi_d + f_s \chi_s$, where $f_d$ and $f_s$ are the fractions of
$B^0_d$ and $B^0_s$ hadrons, respectively, in the sample of semi-leptonic
$B$ decays, and $\chi_f$ is the time-integrated mixing probability for
$B^0_f$.  The quantity $2\cb (1-\cb)$ is the fraction of $b\bar b$ pairs
that decay as like-sign $b$'s.  Our SUSY mechanism can be incorporated by
introducing $\cbe$ such that $2\cbe (1-\cbe)=[2\cb (1-\cb) + G/2]/(1+G)$,
where $G$ is the ratio of SUSY and QCD bottom-quark cross sections after
cuts.  The effective mixing parameter constrains $G$:
\begin{equation}
\cbe=\frac{\cb}{\sqrt{1+G}} +{1\over 2}\left[1-
	\frac{1}{\sqrt{1+G}}\right] \;.
\end{equation}

To estimate $\cbe$, we assume that the world average value $\cb = 0.118 \pm
0.005$ \cite{pdg} represents the contribution from only the pure QCD
component.  We determine the ratio $G$ in the region of phase space where
the measurement is made~\cite{cdfmix}, with both final $b$'s having $p_T$
of at least 6.5 GeV and rapidity $| y_b | \leq 1$.  The ratio is computed
with LO matrix elements, NLO PDF's, $\alpha_s$ at two loops, and $\mu=m_x$,
where $x$ is $b$ or $\tilde g$.  The SUSY and QCD cross sections are
multiplied by 1.9 and 2.3, respectively, to account for NLO effects
\cite{NLOQCD,prospino}.  For gluino masses of $m_{\tilde g} =$ 14 and 16
GeV, we obtain $G =$ 0.37 and 0.28, respectively, with $m_{\tilde b} =$ 3.5
GeV.  We compute $\cbe = 0.17 $ for $m_{\tilde g} =$ 14 GeV, and $\cbe =
0.16 $ with $m_{\tilde g} =$ 16 GeV.

To estimate the uncertainty on $G$, we vary the scale at which the cross
sections are evaluated between $\mu=m_x/2$ and $\mu=2 m_x$.  Uncertainties
of $\pm 50\%$ are obtained and lead to uncertainties in the determined
values of $\delta\cbe\simeq\pm 0.02$.  Additional uncertainties arise
because there is no fully differential NLO calculation of gluino production
and subsequent decay to $b$'s.

Our expectations may be compared with the CDF Collaboration's published
value $\cbe = 0.131 \pm 0.02 \pm 0.016$~\cite{cdfmix}.  Values of
$m_{\tilde g} > 12$ GeV lead to a calculated $\cbe$ that is consistent with
the measured value within experimental and theoretical uncertainties.

The effects of light gluinos and bottom squarks on the strong gauge
coupling $\alpha_s$ are potentially very significant.  In the SM, a global
fit to all observables provides an indirect measurement of $\alpha_s$ at
the scale of the $Z$ boson mass $M_Z$.  The value $\alpha_s(M_Z) \simeq
0.119 \pm 0.006$ describes most observables properly~\cite{alfs}.  QCD
induced processes such as jet cross sections at hadron colliders and the
top-quark cross section are consistent with this value of $\alpha_s(M_Z)$.
A light $\tilde g$ with mass about 15 GeV and a light $\tilde b$ modify
$\alpha_s(M_Z)$, determined by extrapolation from experiments performed at
energies lower than $m_{\tilde g}$.  The result is a shift of 0.007 in
$\alpha_s$ derived from these experiments, to $\alpha_s(M_Z)=0.125$.  The
value of $\alpha_s(M_Z)$ obtained from the $Z$ hadronic width is modified
by the presence of light $\tilde b$'s and light $\tilde g$'s, but the
effect is slight because these superpartners approximately decouple from
the $Z$.  In the presence of light $\tilde g$'s, all values of
$\alpha_s(M_Z)$ still fall within the range of experimental uncertainty,
with a slight preference for the upper edge of the range. We do not attempt
a global fit to the data.

Small values of $m_{\tilde{g}}$ and of the masses of the third generation
squarks are helpful in supersymmetric solutions of the hierarchy problem of
the SM.  They reduce the fine-tuning needed, and avoid a color-breaking
vacuum, in theories where SUSY is broken at the Planck
scale~\cite{dreiner,Gordy}.  Light gluinos are also helpful in improving
gauge coupling unification by providing a light threshold in the evolution
of $\alpha_s$~\cite{unif}.  For gluinos of mass several hundred GeV, the
strong coupling at $M_Z$ predicted from unification is somewhat above 0.13.
However, for gluinos of mass 15 GeV, this prediction becomes $\alpha_s
(M_Z) \alt 0.127$, in agreement with the measured value.  The upper limit
is saturated when the masses of the weak gauginos are of the order of the
weak scale.

Among the predictions of this SUSY scenario, the most clearcut is pair
production of like-sign charged $B$ mesons at the Tevatron collider.  A
very precise measurement of $\cb$ in run II is obviously desirable.  Since
the fraction of $b$'s from gluinos changes with $p_{Tb}$, we also expect a
change of $\cb$ with the cut on $p_{Tb}$.  Possible bound states of bottom
squark pairs are not expected to be prominent in $e^+ e^-$
annihilation~\cite{Nappi} but could be seen as mesonic resonances in
$\gamma \gamma$ reactions or as narrow states in $\mu$ pair production in
hadron collisions.  The existence of light $\tilde b$'s means that they
will be pair produced in partonic processes, leading to a slight increase
in the hadronic dijet rate.  Our approach increases the $b$ production rate
at DESY $ep$ collider HERA and in photon-photon collisions at CERN Large
Electron-Positron Collider (LEP) by a small amount, not enough perhaps if
early experimental indications in these cases are
confirmed~\cite{herab,lepb}, but a full NLO study should be undertaken.
Finally, to satisfy constraints from electroweak measurements, a light
superpartner of the top quark, with mass about the top-quark mass should be
present in the spectrum~\cite{CHWW}.  The lightest Higgs boson should be
lighter than 123 GeV, and it should decay mainly into light $\tilde b$'s.
A study of possible $\tilde b$ decay modes could be undertaken to see if
our scenario can be made consistent with the Higgs-candidate events at
LEP~\cite{LEPH}.

In our investigation, we consider and discard alternative scenarios.  If
only the $\tilde b$ is light, with $\tilde b$ decay products similar enough
to those of the $b$ quark, one can imagine that a light $\tilde b$ might be
sufficient for a good description of the Tevatron $b$ cross section.  The
cross section for pair production of light $\tilde b$'s becomes comparable
to the bottom-quark rate for $m_{\tilde b} \simeq 3$ GeV.  A $\tilde b$
with mass of about 3 GeV decaying through an $R$-parity violating process
into a $\tau$ lepton and a charm quark could lead to an excess of the
apparent $b$ cross section.  This scheme proves difficult to implement.
Such a light $\tilde b$ cannot decay to a $J/\psi$ and will therefore not
reproduce the observed excess $b$ cross section in this decay channel.
Moreover, light $\tilde b$'s do not reproduce the observed $p_T$ dependence
of the $b$ cross section.

If $m_{\tilde b} + m_b > m_{\tilde g}$, but $m_b < m_{\tilde{b}} <
m_{\tilde{g}}$, the gluino can decay into a $\tilde b$ and a strange or
down quark, followed by decay of the $\tilde b$ into a $b$ and a light
neutralino $\tilde {\chi}^0$.  For this scenario, there must be a flavor
violating coupling of \mbox{$\tilde g$ -- $\tilde b$ -- $s$} of about
$10^{-3}$ to suppress the branching ratio of the alternative $\tilde g$
decay channel: $\tilde{g} \rightarrow g \tilde {\chi}^0$.  The problem is
that a light $\tilde b$ proceeding from $\tilde g$ decay, and decaying into
a $b$ and missing energy, would lead to a $b \bar{b}$ plus missing energy
cross section of order of 100~pb, so large that it should have been
observed in the data from run I of the Tevatron collider~\cite{missE}.

In this Letter, we propose an interpretation of the excess bottom-quark
production rate at the Tevatron that involves new physics and leads to
several testable consequences.  We show that pair production of low-mass
gluinos that decay into bottom quarks and bottom squarks provides a good
description of the $b$ cross section and reproduces the measured ratio of
like-sign to opposite-sign leptons.  Our assumptions are consistent with
all experimental constraints on SUSY particle masses.

{\bf Acknowledgements} This work is supported by the U.S. Department of
Energy under Contract W-31-109-ENG-38.  We thank the Aspen Center for
Physics and the University of California, Santa Cruz, where part of this
work was done, and K.~Byrum, M.~Carena, H.~Frisch, H.~Haber, T.~LeCompte,
S.~Mrenna, A.~Nelson, D.~Wackeroth, A.~B.~Wicklund, J.~Womersley, and
D.~Zeppenfeld for insightful physics comments.


\end{document}